\documentclass[journal=jacsat,manuscript=article]{achemso}

\usepackage{chemformula} % Formula subscripts using \ch{}
\usepackage[T1]{fontenc} % Use modern font encodings
\usepackage{multicol}
\usepackage{graphicx}

\usepackage{float}
\usepackage[verbose]{placeins}
\author{Abhimanyu Singareddy}
\affiliation{Department of Electrical Engineering, Indian Institute of Technology Bombay, Mumbai, 400076, India}

\email{abhimanyumatrix2@gmail.com}

\author{Pradeep R. Nair}
\affiliation{Department of Electrical Engineering, Indian Institute of Technology Bombay, Mumbai, 400076, India}
\email{prnair@ee.iitb.ac.in}

\title{Influence of Phase Segregation on the Hysteresis of Perovskite Solar Cells}

\begin{document}
\begin{abstract}
Organic-inorganic hybrid perovskite solar cells (PSC) have demonstrated impressive performance improvement. Among the various characteristics, the time-dependent current-voltage (J-V) hysteresis allows a direct exploration of various critical phenomena that affect the stability of PSCs. The hysteresis is associated with various spatial heterogeneity-related phenomena, including lifetime, bandgap, and phase segregation. We investigate these phenomena through numerical simulations and quantify how the spatial non-uniformity in the perovskite active layer impacts the hysteresis. Further, we correlate the time dependent device degradation with the hysteresis trends in terms of ion density and effective carrier lifetime.
 
\end{abstract}

\maketitle

\section{I. Introduction}% 1
Perovskite based solar cells (both single junction and tandem) are increasingly considered as a viable substitute or value addition to the conventional silicon technology due to their cost-effectiveness and ease of fabrication \cite{Jena2019}. The versatility in tuning the bandgap of mixed halide perovskites offers a distinct advantage for applications such as LEDs \cite{Zhang2020}, subcells in tandem solar cells\cite{Bush2017}, etc. These mixed halide perovskites are unstable under illumination, leading to a phenomenon called phase segregation (PS) \cite{Hoke2015a}. Due to PS, performance parameters such as open circuit voltage, short circuit current, fill factor, and efficiency are critically affected \cite{Samu2017a}. Numerous investigations \cite{Li2017, Brennan2018b} revealed that the PS enhances ion migration and charge carrier trap creation, further contributing to the hysteresis. This could lead to inaccurate estimates for the performance metrics \cite{Tress2015}. \\
Given the excellent improvement in power conversion efficiency, stability and degradation have emerged as the critical aspects that might dictate the commercial viability of perovskite based solar cells. Among the several phenomena, ion migration and interface degradation have a direct influence on the time dependent efficiency degradation of such PSCs. There are only very few experimental techniques that allow direct and simple quantification of these critical phenomena. One of them is the transient hysteresis measurement \cite{VanReenen2015, Snaith2014, Habisreutinger2018a, Chen2016}. Here, hysteresis denotes the difference between the current-voltage (J-V) curves (see Section A in the Supplementary Material for more details) obtained by sweeping the voltage from the short circuit to the open circuit condition (forward scan) and back (reverse scan) \cite{Habisreutinger2018a}. Many factors \cite{Lin2022} influence hysteresis, such as the poling voltage, scan rate, scan direction, ion densities, the structure of the device, ion mobilities, etc. Although the hysteresis is secondary to steady-state efficiency, these measurements still offer good insights into the state of the device in terms of ion migration, the extent of PS, and the location of high recombination regions in the perovskite layer. \\
Originally, several mechanisms \cite{Chen2016, Liu2019} have been proposed to explain hysteresis, such as ion migration, ferroelectric polarization, charge trapping/de-trapping, etc. Different models such as drift-diffusion based on ion migration\cite{VanReenen2015}, dynamic electric model\cite{Nemnes2019}, and polarization model based on charge accumulation \cite{Ravishankar2017} have been developed to understand the hysteresis behavior. Through simulations, Reenen et al. \cite{VanReenen2015} demonstrated that hysteresis is caused by ion migration and charge carrier recombination. However, in addition to interface effects, significant other spatial non-homogeneities could contribute to hysteresis. For example, grain boundaries could trap charges and contribute to localized carrier recombination \cite{Sherkar2017a, Nandal2019}. In this context, the recent reports indicate that crystal grain sizes influence hysteresis \cite{Cao2016a, Kim2014}, but the influence of the defect location (e.g.: grain boundary) on hysteresis is not explored. Li et al. \cite{Li2017} observed that the PS phenomenon generates mobile ions, that could migrate to the perovskite/transport layer interface resulting in injection barriers and thus leading to J-V hysteresis. However, the influence of the geometry or PS induced spatial heterogenity on the hysteresis in PSCs is yet to be explored. \\    
In this manuscript, we explore the influence of spatial heterogeneties on the hysteresis of the PSC. These effects could be localized or distributed in the active region, resulting in spatial non-uniformity that impacts the effective carrier lifetime ($\tau_{eff}$) of the system (see Section B of the supplementary material for more details). Examples of such spatial non-uniformities are increased recombination at the grain boundaries and/or multiple domains due to phase segregation. Of these two, addressing the influence of phase segregation is a challenging task because of the complexity associated with numerical simulations. In our previous work \cite{Singareddy2021}, we elucidated the impact of the PS phenomenon on the device's steady-state efficiency through statistical simulations by considering different spatial patterns, incorporating ion migration and increased non-radiative recombination. Such phase segregated domains could act as potential wells (due to the band level mismatch between adjacent domains) for carriers thus leading to increased carrier recombination. There exist no quantitative estimates for the influence of such segregated domains on the hysteresis. Further, an interesting prospect is the possibility of quantifying the extent of phase segregation from hysteresis measurements.\\ 
In the next section, we discuss the model system used to simulate time dependent J-V characteristics of PSC under phase segregation. Later, we investigate various spatial heterogeneity effects in the perovskite layer and their influence on the device's hysteresis. Finally,  we correlate the degradation in the active layer over time in terms of carrier lifetime and ion densities.

\section{II. Simulation methodology and model system}
\begin{figure*}[!htbp]
	\centering
	\includegraphics[scale = 0.6]{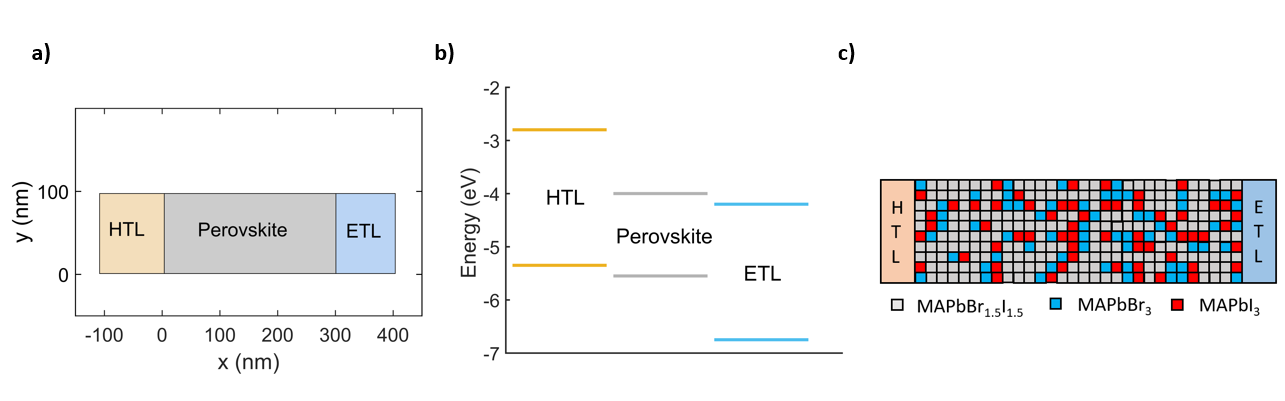}
	\caption{
		a) Schematic of a pristine 2D device structure consisting of hole transport layer (HTL), perovskite active layer, and electron transport layer (ETL) used for simulating hysteresis, and b) energy band alignment of layers, and c) schematic of the device model used to study the effect of phase segregation on the hysteresis of the device.
	}
	
\label{fig1}	
\end{figure*}

\begin{figure*}[!htbp]
	\centering
	\includegraphics[scale = 0.55]{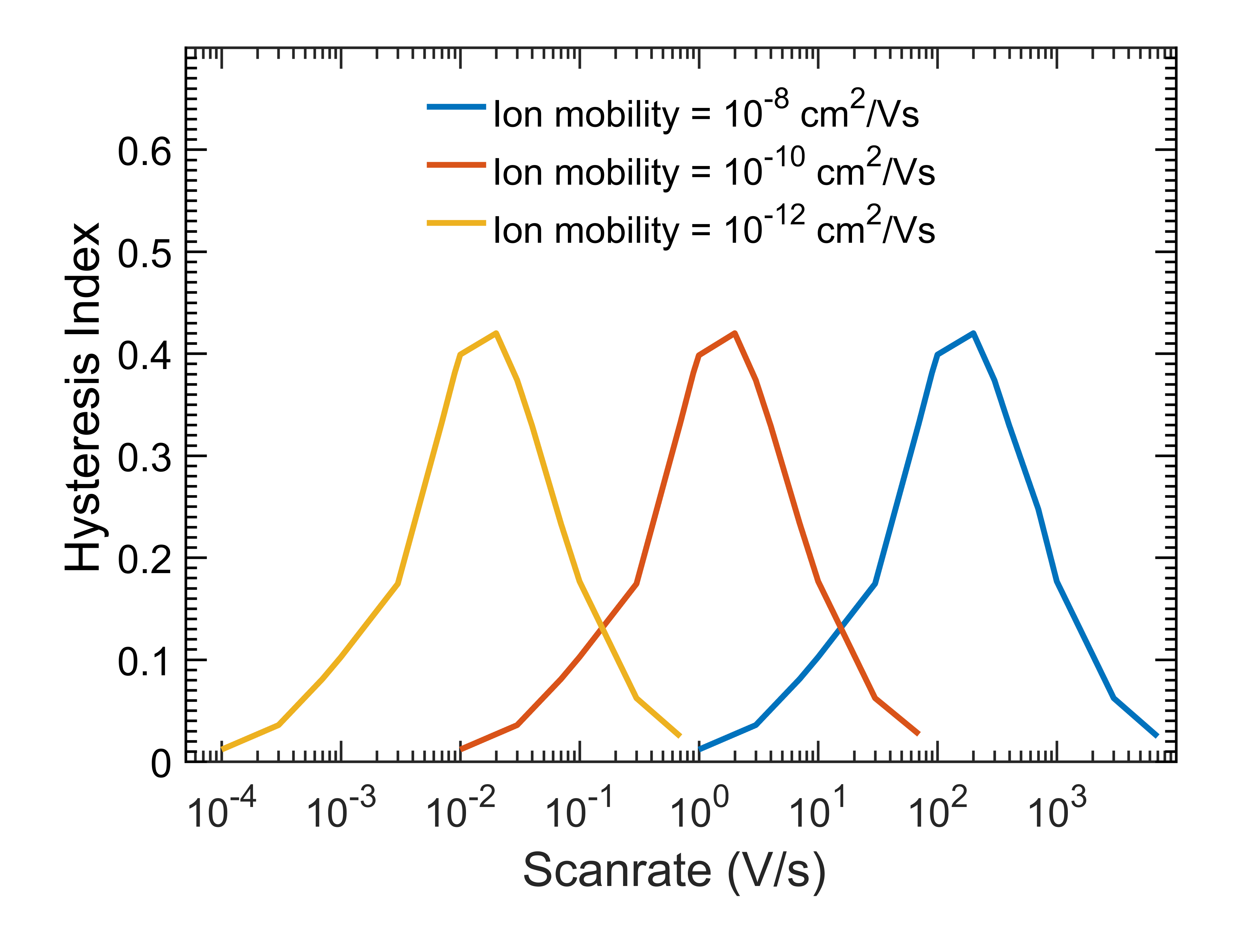}
	\caption{
		 HI vs. scan rate for different ion mobilities plotted for an ion density of $ 5\times10^{17}$ cm$^{-3}$. An effective carrier lifetime of $10$ ns is considered for generating the curves.
	}
\label{fig2}	
\end{figure*}

The schematic of the perovskite device (PIN structure) is shown in Fig. \ref{fig1}a. The energy level alignments of the pristine sample is shown in Fig. \ref{fig1}b.
The device consists of a hole transport layer (HTL) with an acceptor doping density and an electron transport layer (ETL) with a donor doping density. In accordance with literature \cite{Neukom2019}, the perovskite active layer is assumed to have immobile positive and mobile negative ions (the net ion density is zero).  We assume an injection barrier (from transport layers to the perovskite layer) of 0.2 eV on both sides of the perovskite layer (see Table S1 in the supplementary material for numerical values). For the mentioned model system, Poisson’s equation for electrostatics (see Eq. 1), time-dependent continuity equations for holes, electrons, and mobile ions (see Eq. 2-4), and transport equations (see Eq. 5-7) are solved self consistently. These equations are given by, \\

\begin{equation}
	\nabla. \nabla\phi = -\frac{q}{\varepsilon \varepsilon_0}(p-n+\rho_I+N_D-N_A),
	\end{equation}
		\begin{equation}
	\frac{\partial n}{\partial t} = \frac{1}{q} \nabla J_n+G-R,
	\end{equation}
		\begin{equation}
	\frac{\partial p}{\partial t} = -\frac{1}{q} \nabla J_p+G-R
	\end{equation}
	\begin{equation}
		\frac{\partial N_{I,n}}{\partial t}= \frac{1}{q} \nabla J_{I,n},
	\end{equation}
	\begin{equation}
	J_n = q\mu_n nE+qD_n \nabla n,
	\end{equation}
	\begin{equation}
	J_p = q\mu_p pE-qD_p \nabla p,
	\end{equation}
	\begin{equation}
	J_{N_{I,n}}= q\mu_{N_{I,n}} N_{I,n}E+qD_{N_{I,n}} \nabla N_{I,n},
	\end{equation}
where $\phi$ is the electrostatic potential, $\varepsilon, \varepsilon_0$ are relative and absolute permittivity of the material, p and n are hole and electron densities, $\rho_I$ is the net ionic density (i.e., $\rho_I= N_{I,p}-N_{I,n}$,  where $N_{I,p}$  and  $N_{I,n}$ are immobile positive and mobile negative ionic charge densities, respectively), and $N_D$, $N_A$ are donor and acceptor doping densities respectively. The terms G and R denote the generation and recombination rates for the electrons/holes, respectively. Recombination (R) of charge carriers is modeled (see Section B of the supplementary material for detailed information) by using the radiative, Auger, and Shockley-Read-Hall (SRH) mechanisms. Interface recombination (transport layer/perovskite interface) is accounted through appropriate recombination velocities. Also, the terms $\mu_n$, $\mu_p$, $\mu_{N_{I,n}}$ represent the mobilities, and the terms $D_n,D_p ,D_{N_{I,n}}$ are the diffusion coefficients of electrons, holes, and mobile ions respectively. Note that the above set of equations (Eq. 1-7) were previously used \cite{Neukom2019,SakethChandra2021, Nandal2017} to explore the device physics of PSCs. The parameters used for the simulations are listed in Table S1 of the supplementary material. \\
The mixed halide perovskites are known to undergo phase segregation upon illumination \cite{Hoke2015a,Chai2021}. Here, we consider mixed perovskite (MAPbBr$_{1.5}$I$_{1.5}$) as model system that segregates into pure iodide (MAPbI$_{3}$) and bromide (MAPbBr$_{3}$) domains at random locations upon illumination. Figure \ref{fig1}c shows the schematic of the phase segregated device. To simulate the random PS phenomenon, we use the following methodology \cite{Singareddy2021}: the entire perovskite material is treated as a collection of smaller domains, where the properties of each domain are specified individually. These domains can be either iodide, bromide, or mixed halide. The placement of domains is as follows: a bromide domain is placed randomly, and a corresponding iodide domain is set within the neighborhood of bromide. We repeat this process till we achieve the required fraction of PS. Finally, the remaining domains are assigned the properties of mixed halide (see the schematic in Fig. \ref{fig1}c). The effect of phase segregation on steady-state efficiency, explored through numerical simulations under steady state Boltzmann statistics for ions, was reported in our previous work \cite{Singareddy2021}. In this manuscript, we address the effect of phase segregation on the hysteresis of the device - i.e., scan rate dependent JV characteristics of the device (which is computationally more complex than the steady state simulations). \\
To quantify the degree of hysteresis in the device \cite{Habisreutinger2018a}, we use the term hysteresis index (HI) given by,
\begin{equation}\label{eq:4}
HI=  \frac{PCE(reverse)-PCE(forward)}{PCE(reverse)},
    \end{equation}
where PCE (reverse) is the power conversion efficiency in the reverse scan and PCE (forward) is the power conversion efficiency in the forward scan. \\
   \FloatBarrier
 From literature \cite{courtier2019modelling}, it is well known that a unique scan rate produces a maximum hysteresis for a given set of parameters. Figure \ref{fig2}  shows the HI vs. scan rates for different cases of ion mobilities for a simple PIN device. The peak hysteresis is observed at a particular scan rate. Above this scan rate, the slow-moving ions do not respond to the fast changes in the external voltages. On the contrary i.e., below the specified scan rate, ions have ample time to reach a steady state, which reduces the difference between the forward and reverse curves. For the specified scan rate, ions respond to the changes in the applied voltage and redistribute according to the electric field inside the perovskite, thereby controlling the recombination rate and the current flow. These results are in accordance with literature \cite{Huang2022} and indicate  that the ion mobilities and scan rates play a crucial role in determining the HI (see Fig. \ref{fig2}). 
          \FloatBarrier
A range of ion mobilities are reported in the literature \cite{Lee2019} i.e., from $10^{-8}$ cm$^{2}$/Vs to $10^{-14}$ cm$^{2}$/Vs. As discussed earlier, for each of these ion mobilities, there will be a unique scan rate where maximum hysteresis is observed. For ion mobility of $10^{-10}$ cm$^{2}$/Vs and uniform ion density of $ 10^{17}$ cm$^{-3}$, the scan rate at which maximum hysteresis obtained is 0.2 V/s (see Fig. S2 in the supplementary material).  Similarly, for ion mobility of $10^{-10}$ cm$^{2}$/Vs and uniform ion density of the order of $ 5\times10^{17}$ cm$^{-3}$ the scan rate at which maximum hysteresis obtained is 2 V/s (see Fig. \ref{fig2}). These values, which are similar to other reports in the literature \cite{Huang2022}, are used (unless specified) to simulate the different phenomena discussed in this manuscript. \\
The main objective of this manuscript is to address the PS effect on the time-dependent current-voltage hysteresis of PSCs. PS could lead to a spatial distribution of defects and domains with different band gaps. Numerical simulation of such phase segregated heterojunction structures involving complex spatial patterns is computationally complex.  To assimilate insights in an incremental manner, we start with simple cases to explore different heterogeneity scenarios and examine their impact on the hysteresis of the device. Such accumulated insights aid to comprehend the physics behind the hysteresis in a phase segregated device. Accordingly, we first consider the influence of localized defects in the active layer on hysteresis. We then consider the influence of individual localized domains of lower band gap in the active layer. Such domains could act as potential wells for carriers and enhance the recombination thus contributing to hysteresis. Finally, we consider the influence of phase segregation with multiple domains on the hysteresis.
\section{III. Influence of localized recombination centers}
One of the critical phenomena affecting perovskite's performance is the localized recombination centers/defects that arise due to phase segregation or grain boundaries in the perovskite. These defects can create or act as trapping or recombination centers within the bandgap of the perovskite material \cite{Ni2021, Biaou2024}. These trap centers reduce the effective lifetime of the device, leading to the capture of carriers from their respective bands or/and the recombination of charge carriers. Cao et al. \cite{Cao2016a} experimentally proved that HI is influenced by grain sizes, i.e., large grain sizes will have fewer grain boundaries and less recombination, and vice versa. Therefore, these defects can influence the performance depending on the position where they are created in the active layer.
\FloatBarrier
   \begin{figure*}[!htbp]
	\centering
	\includegraphics[scale = 0.8]{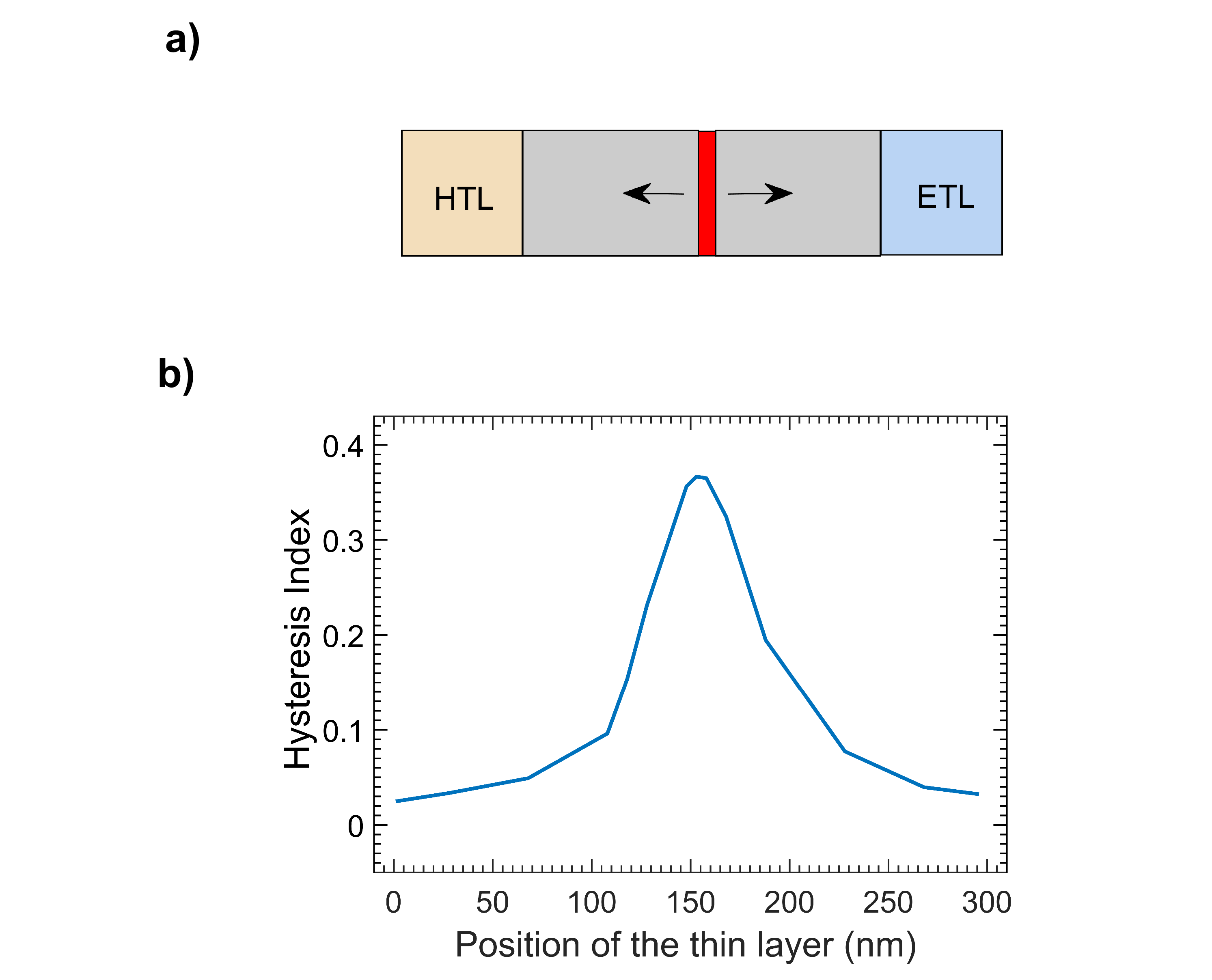}
	\caption{
		a) Schematic of the model system used to study the effect of thin recombination layer denoted by red in the active region. b) HI vs. the position of the thin recombination layer varied across the thickness of the active layer.
	}
	
\label{fig3}	
\end{figure*}
To study the effect of the spatial location of defects on the hysteresis, we introduce a thin layer with thickness of 5 nm (denoted by red color in Fig. \ref{fig3}a) in the active layer ($\tau_{eff} =100$ ns) with the same properties as the active layer except for a low effective carrier lifetime ($\tau_{eff}=1$ ns). Due to the low carrier lifetime in the thin layer, the significant charge carrier recombination happen in this layer.  The layer could be a grain boundary \cite{Nandal2019} or a highly degraded region of perovskite. This thin layer is placed perpendicular to the current flow (see Fig. \ref{fig3}a) and varied along the thickness of the active region.\\
 The schematic of the device with the thin recombination layer and its effect on the hysteresis is shown in Fig. \ref{fig3}. HI reaches its peak when the position of the thin defective layer is at the center. HI reduces from its peak value as the recombination layer moves towards the HTL/ETL interface. It is well known that the carrier recombination rate is at its maximum, where both electrons and holes are large and equal in concentration (i.e., n $=$ p). In this case, it is at the center of the active layer (see Fig. S3 in the supplementary material). This recombination is higher in the forward scan than in the reverse scan (see Section A in the supplementary material for more details) resulting in higher HI. As the band offsets are symmetrical, HI is expected to be symmetrical at either side and to be maximum precisely at the mid-position (in this case, it is at 150 nm across a 300 nm thickness active layer) of the perovskite layer. However, the ion distribution is asymmetrical at the interfaces because of positive immobile ions and negative mobile ions (see Fig. S4 in the supplementary material). This leads to the asymmetrical distribution of charge carriers (see Fig. S3 in the supplementary material). Therefore, HI is slightly larger towards the ETL interface than HTL interface.  \\
  From Fig. \ref{fig3}b, we conclude that the hysteresis in PSC is sensitive to position-dependent recombination centers i.e., the low lifetime region/defects or grain boundary at the middle region of the active layer has the maximum impact on the J-V hysteresis of the solar cell. 
\section{IV. Influence of localized low bandgap domain}
In this section, we address the influence of localized low bandgap domain in the perovskite layer on the hysteresis in a solar cell. To simulate the effect, we introduce a low bandgap material of 10 nm thickness positioned perpendicular to the current flow. The location of this low band gap domain is varied along the thickness of the active layer. The proposed scheme is similar to the scheme presented in Section III  where a thin layer with a low lifetime is introduced. \\
We introduce a pure iodide domain (MAPbI$_3$) in the mixed halide active layer (MAPbBr$_{x}$I$_{1-x}$) to simulate the effect of low bandgap material on hysteresis. The energy levels of different materials are plotted in Fig. S5 in the supplementary material. The generation rates for the materials used are as per earlier reports \cite{Singareddy2021} i.e., for low bandgap iodide domain $G_{I}=4.53\times 10^{21}$ cm$^{-3}$s$^{-1}$ and for the mixed halide domain $ G_M=2.98\times 10^{21}$ cm$^{-3}$s$^{-1}$. Some aspects can be anticipated through careful examination of band level alignment. The conduction band offset of $\approx$ 0.3 eV between the mixed halide domain and iodide domain (see Fig. S5 in the supplementary material) acts as a potential well and prevents most of the electrons in the iodide domain from undergoing over the barrier transport, whereas the holes in the iodide domain observe a negligible offset. Due to the high generation rate, low bandgap, and favorable energy levels, the recombination in the device is dictated by the position of the low bandgap iodide domain.
\FloatBarrier
\begin{figure*}[!htbp]
  \centering
   \includegraphics[width=0.5\textwidth]{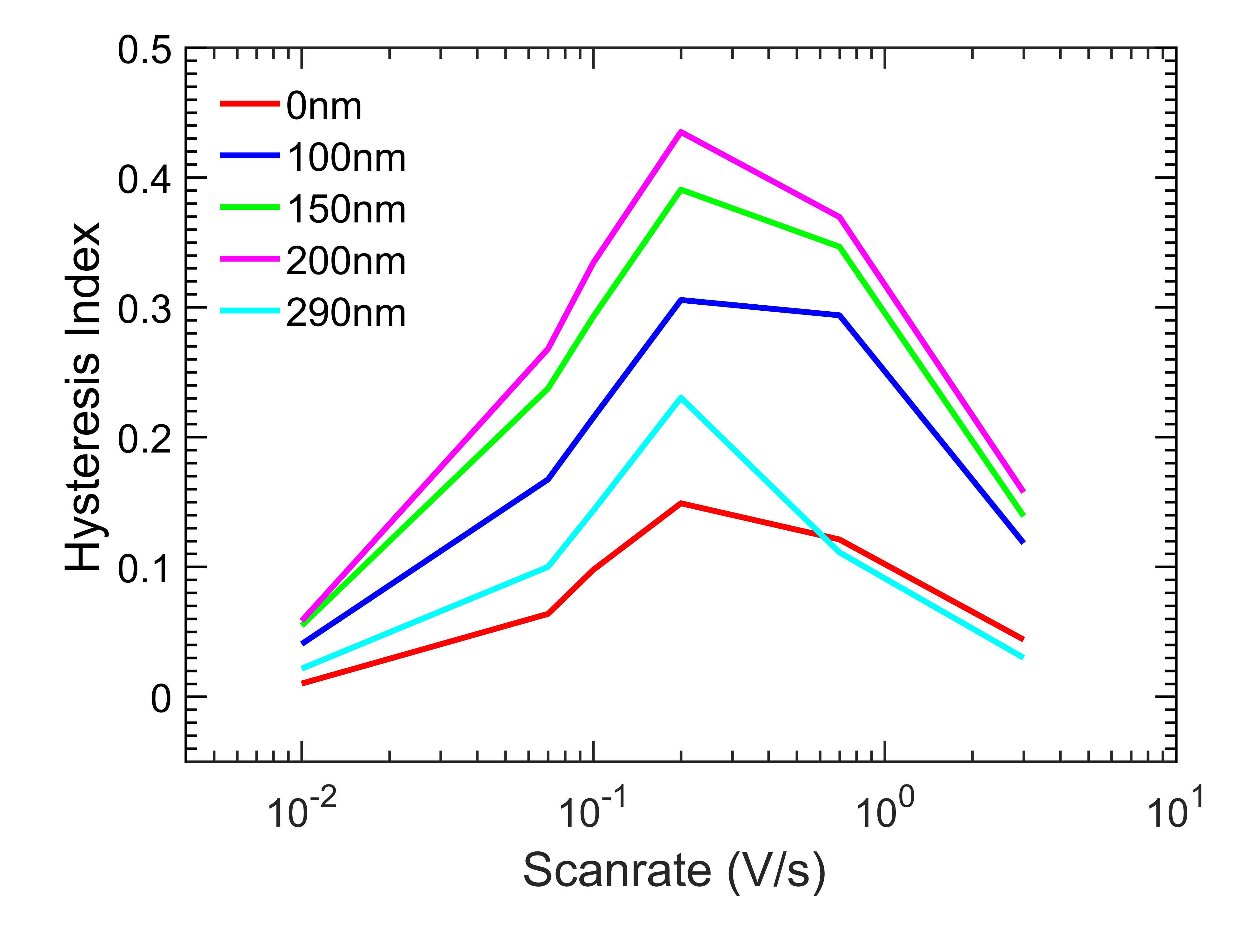}
      \caption{Hysteresis index (HI) vs. scan rate for different positions of the low bandgap domain varied across the thickness of the active layer.}
\label{fig4}
%\raggedbottom
\end{figure*}
The effect of the position of the low bandgap domain in the active layer on hysteresis is shown in Fig. \ref{fig4}. The curves are similar to the trends seen in Fig. \ref{fig3}b. The lower HI is seen when the low bandgap iodide domain is at the 0 nm position (i.e., near HTL interface) and the 290 nm position (i.e., near ETL interface). The HI is high when the low bandgap iodide domain is positioned at 200 nm because of the higher densities of electrons and holes (see Fig. S6 in the supplementary material) than at the precise mid position. The above analysis shows that the HI is sensitive to the position of the low band gap domain in the mixed perovskite active layer. These insights allow us to explore the hysteresis in a phase-segregated device, as detailed below.
\section{V. Influence of phase segregation}
\begin{figure*}[!htbp]
	\centering
	\includegraphics[scale = 1.2]{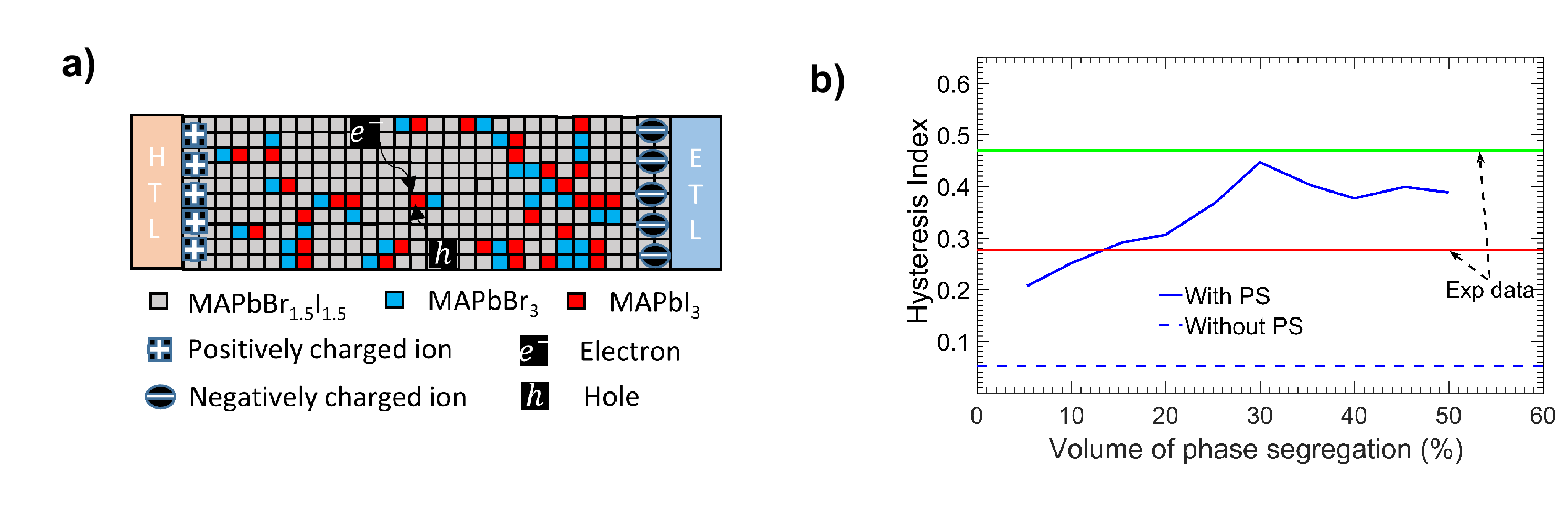}
	\caption{
		a) Schematic of 2D model system with 300 domains in the active layer used to study hysteresis in an ion populated, phase segregated mixed halide perovskite device. The schematic shows the recombination of holes and electrons in the low bandgap iodide-rich domain. b) HI is plotted for different levels of PS in the active layer. The solid blue line denotes HI with PS, and the dashed blue line denotes HI without PS. Black arrows refer to experimental HI data from literature: green lines denote data from ref  \cite{Li2017} and red lines denote data from ref  \cite{Chai2021}.
	}
\label{fig5}	
\end{figure*}

Here we consider a phase segregated device with multiple domains of different materials and heterojunctions between them (see Fig. \ref{fig5}a). To simulate the hysteresis, the phase segregated device with mobile ions (at an initial uniform density of $ 1\times10^{17}$ cm$^{-3}$) is considered. For ion mobility of $10^{-10}$ cm$^{2}$/Vs, the scan rate at which maximum hysteresis occurs is 0.2 V/s (see Fig. S2 in the supplementary material for more details). We used $G_{I}=4.53\times 10^{21}$ cm$^{-3}$s$^{-1}$, $G_B=1.48\times 10^{21}$ cm$^{-3}$s$^{-1}$, and $ G_M=2.98\times 10^{21}$ cm$^{-3}$s$^{-1}$ as the uniform photo-carrier generation rates for the iodide, mixed halide, and bromide domains, respectively \cite{Singareddy2021}. The energy levels of these materials are plotted in Fig. S7 of the supplementary material.
 \FloatBarrier
The HI data for different volumes of PS is plotted in Fig. \ref{fig5}b. We notice a steady increase in HI with PS. With the increase in PS, the density of high bandgap bromide and low bandgap iodide domains increases. The recombination in the low bandgap iodide domain increases depending on their location (see Section IV for more details) resulting in high HI. These high HI values can affect the reliability of the measurement data by under/overestimating the performance parameters \cite{Tress2015}. The J-V characteristics of the phase segregated PSC under illumination in forward and reverse scans for different volumes of PS are plotted in Fig. S8 of the supplementary material.\\
   \textbf{Comparison with experimental data:}
   The simulated HI values are comparable with available experimental data \cite{Li2017, Chai2021}. Chai et al. \cite{Chai2021} observed PS in the CsPbIBr$_2$ ﬁlm under illumination. The J-V measurement in forward as well as reverse scans yielded a high HI of 0.277 for the phase segregated device. An effective strategy was suggested to arrest the PS in CsPbIBr$_2$ films by improving their crystalline grains with PMMA resulting in a low HI of 0.058 \cite{Chai2021}. This is similar to the analysis of localized recombination centers described in Section III. Similarly, Li et al. \cite{Li2017} observed PS-generated iodide domains at grain boundaries in the CsPbIBr$_2$ ﬁlm. Along these grain boundaries, PS generates a high density of mobile ions that could create injection barriers by piling at the perovskite/TiO$_2$ interface resulting in a high HI of 0.47. The HI is reduced to 0.066 by modifying the interface of perovskite and the transport layer of a phase segregated device \cite{Li2017}. Although the HI values were reported \cite{Li2017, Chai2021}, it is not easy to experimentally determine the level of phase segregation. Nevertheless, our numerical simulations and the experimental data demonstrate that HI increases in a phase segregated device compared to the device without PS (see Fig. \ref{fig5}b). The hysteresis is affected by PS as well as the migration of the ions to the interface of the perovskite and the transport layer. Therefore, modifying the interface and arresting PS is crucial to reduce HI. This work quantitatively relates the HI to the volume of phase segregation in mixed halide perovskite systems.

\section{VI. Estimation of degradation in the device}
\begin{figure*}[!htbp]
	\centering
	\includegraphics[scale = 0.55]{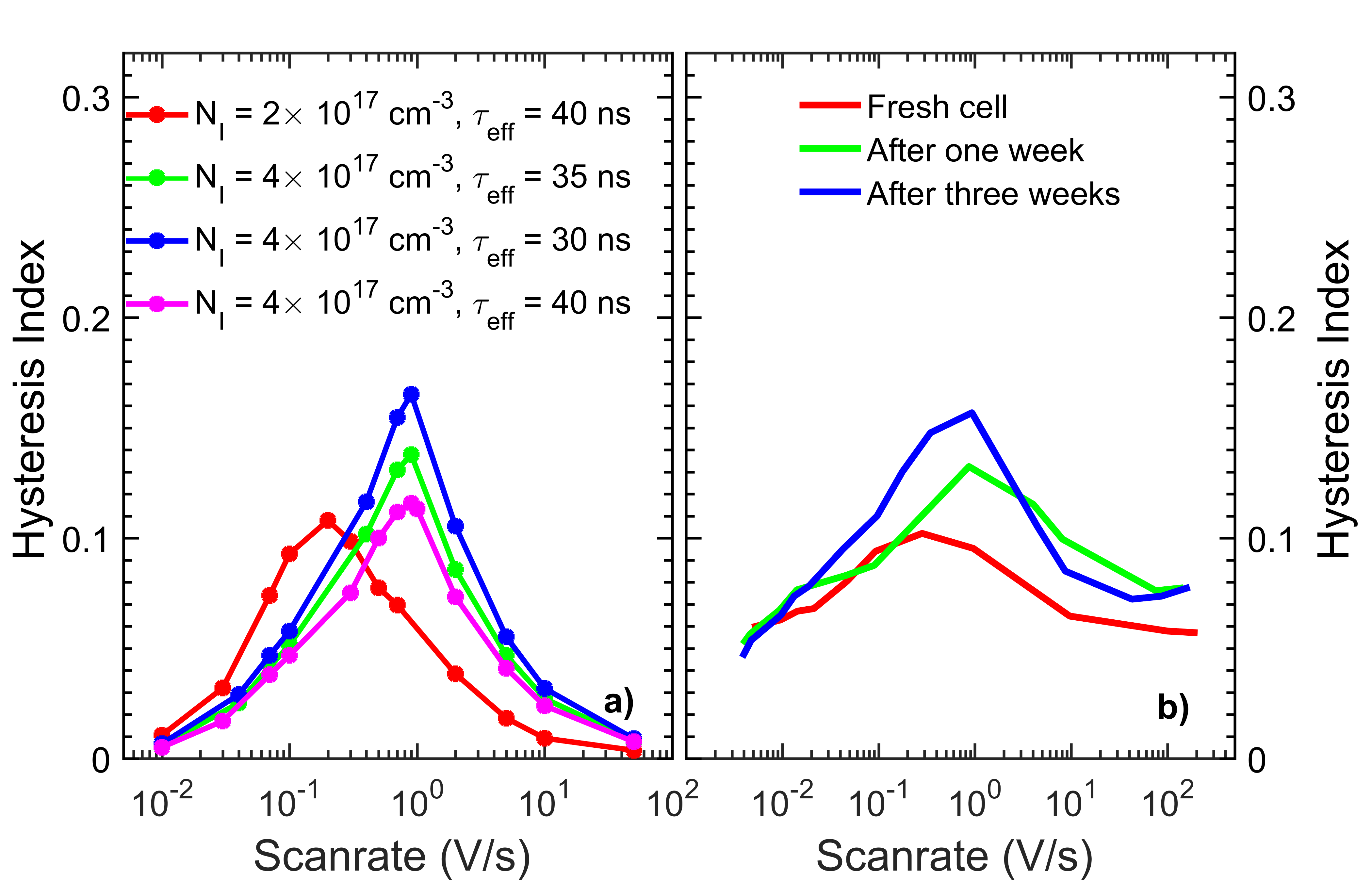}
	\caption{
		a) Simulated HI vs. scan rate for different ion densities and carrier recombination lifetimes, and b) experimental HI values \cite{Nemnes2019} vs. scan rate for a reference fresh sample, aged for one week, and aged for three weeks.
	}
	
\label{fig6}	
\end{figure*}

The degradation of perovskite increases with illumination time, exposure to oxygen, humidity, etc \cite{Boyd2019}. Provided this, one question arises: "Can we quantify the device degradation in terms of ion generation and carrier lifetime using the hysteresis index?" To answer this, we compare the experimental data in the literature \cite{Nemnes2019} to our numerical simulations. Here, we analyze the hysteresis effect with different ion densities and recombination lifetime. Fig. \ref{fig6}a shows simulated HI vs. scan rates for different ion densities and recombination lifetimes. The HI for various ion densities with the same recombination lifetime is plotted in Fig. S9 in the supplementary material.\\
We simulated the model device with an ion density of $2\times10^{17}$ cm$^{-3}$ and an effective recombination lifetime of 40 ns (see dotted red line in Fig. \ref{fig6}a). The HI peak is seen for a scan rate of 0.2 V/s. When the ion density is increased from $N_{I} = 2\times10^{17}$ cm$^{-3}$ to $N_{I} = 4\times10^{17}$ cm$^{-3}$, the scan rate at which peak HI occurs shifts to the right (see the magenta line in Fig. \ref{fig6}a). We observe that the peak HI shifts to higher scan rates as ion density increases for the same carrier lifetime (compare between red and magenta lines in Fig. \ref{fig6}a). When the effective charge carrier lifetime is reduced, the scan rate at which peak HI occurs remains the same, but the magnitude of HI increases (see Fig. \ref{fig6}a). These insights are useful to analyze the available experimental data.\\
Nemnes et al. \cite{Nemnes2019} reported hysteresis measurements on a fresh cell, after one week and three weeks later (see Fig. \ref{fig6}b). Here, we notice that for the fresh cell, the HI peak value is lower (see red solid line in Fig. \ref{fig6}b) and appears at a different scan rate than the aged cell (blue or green curves). The reason could be that the fresh cell has a lower ion density than the aged cell. With time, the degradation of perovskite occurs due to various reasons \cite{Boyd2019} such as light, heat, moisture, oxygen etc. Accordingly, we expect an increase in ion density and a decrease in overall carrier lifetime due to an increase in defects. As predicted by numerical simulations, the experimental HI shows two distinct features: (a) The peak shifts to higher scan rate and (b) the magnitude of HI increases. These trends could be due to an increase in the density of mobile ions and a decrease in the carrier lifetime. The peaks of experimental and simulated HI curves occur at the same scan rates and have the same value, but there is a difference in the shape of the curves (elevated HI at the extreme scan rates, see Fig. \ref{fig6}b). The possible reasons could be ferroelectric effects \cite{Chen2015a} as well as unbalanced charge transport\cite{Heo2015} - which needs further exploration.
\FloatBarrier
To summarize, we performed numerical simulations to quantify the degradation in the device in terms of ion density and charge carrier recombination lifetime using hysteresis data. This suggests that stability of PSCs could be explored through transient measurements which can simultaneously track the ion generation and lifetime degradation. 
\section{VII. Conclusions}
Through numerical simulations, we investigated how various spatial heterogeneities such as grain boundaries and phase segregation affect J-V hysteresis. Our key results demonstrate that the hysteresis data provide information about the device's state regarding ion generation, lifetime degradation, and PS. This allows to quantify and correlate hysteresis measurements with long-term stability issues under real-time conditions.

\section{VIII. Supplementary Material}
See supplementary material for details regarding J-V hysteresis in a PSC, modeling of the recombination mechanism and parameters, HI vs scan rate plots, ion and charge carrier density plots at equilibrium, material energy levels, and J$—$V characteristics for a phase segregated device.
\section{IX. Acknowledgements}
This work was funded in part by Science and Engineering Research Board (SERB, project code CRG/2019/003163), Department of Science and Technology (DST), India. The authors acknowledge IITBNF and NCPRE for computational facilities. A.S acknowledge the financial support of University Grants Commission (UGC), P.R.N. acknowledges Visvesvaraya Young Faculty Fellowship.

\bibliography{aipsamp}
\end{document}